# The scientific case for brain simulations

Gaute T. Einevoll[1,2,*], Alain Destexhe[3,4], Markus Diesmann[5,6,7], Sonja Grün[5,8], Viktor Jirsa[9], Marc de Kamps[10], Michele Migliore[11], Torbjørn V. Ness[1], Hans E. Plesser[1,5], Felix Schürmann[12]

[1]Faculty of Science and Technology, Norwegian University of Life Sciences, 1432 Ås, Norway
[2]Department of Physics, University of Oslo, 0316 Oslo, Norway
[3]Paris-Saclay Institute of Neuroscience (NeuroPSI), Centre National de la Recherche Scientifique, 91198 Gif-sur-Yvette, France
[4]European Institute for Theoretical Neuroscience, 75012 Paris, France
[5]Institute of Neuroscience and Medicine (INM-6) and Institute for Advanced Simulation (IAS-6) and JARA-Institut Brain Structure-Function Relationships (INM-10), Jülich Research Centre, 52425 Jülich, Germany
[6]Department of Psychiatry, Psychotherapy and Psychosomatics, RWTH Aachen University, 52074 Aachen, Germany
[7]Department of Physics, RWTH Aachen University, 52074 Aachen, Germany
[8]Theoretical Systems Neurobiology, RWTH Aachen University, 52074 Aachen, Germany
[9]Institut de Neurosciences des Systèmes (INS), Inserm, Aix Marseille Univ, 13005 Marseille, France
[10]Institute for Artificial and Biological Intelligence, School of Computing, Leeds LS2 9JT, United Kingdom
[11]Institute of Biophysics, National Research Council, 90146 Palermo, Italy
[12] Blue Brain Project, École polytechnique fédérale de Lausanne (EPFL), Campus Biotech, 1202 Geneva, Switzerland

*Corresponding Author: Gaute T. Einevoll, Faculty of Science and Technology, Norwegian University of Life Sciences, PO Box 5003, 1432 Ås, Norway; Gaute.Einevoll@nmbu.no

**In Brief**

A key element of several large-scale brain research projects such as the EU Human Brain Project is simulation of large networks of neurons. Here it is argued why such simulations are indispensable for bridging the neuron and system levels in the brain.

**Abstract**

A key element of the European Union's Human Brain Project (HBP) and other large-scale brain research projects is simulation of large-scale model networks of neurons. Here we argue why such simulations will likely be indispensable for bridging the scales between the neuron and system levels in the brain, and a set of brain simulators based on neuron models at different levels of biological detail should thus be developed. To allow for systematic refinement of candidate network models by comparison with experiments, the simulations should be multimodal in the sense that they should not only predict action potentials, but also electric, magnetic, and optical signals measured at the population and system levels.

# 1      Introduction

Despite decades of intense research efforts investigating the brain at the molecular, cell, circuit and system levels, the operating principles of the human brain, or any brain, remain largely unknown. Likewise, effective treatments for prevalent serious psychiatric disorders and dementia are still lacking (Hyman, 2012; Masters et al., 2015). In broad terms one could argue that we now have a fairly good understanding of how individual neurons operate and process information, but that the behavior of networks of such neurons is poorly understood. Following the pioneering work of Hubel and Wiesel mapping out receptive fields in the early visual system (Hubel and Wiesel, 1959), similar approaches have been used to explore how different types of sensory input and behaviour are represented in the brain. In these projects the statistical correlation between recorded neural activity, typically action potentials from single neurons, and sensory stimulation or behaviour of the animal is computed. From this, so-called *descriptive* mathematical models have been derived, accounting for, say, how the firing rate of a neuron in the visual system depends on the visual stimulus, see e.g., Dayan and Abbott (2001, Ch. 2).

The qualitative insights gained by obtaining these descriptive receptive-field models should not be underestimated, but these models offer little insight into how networks of neurons give rise to the observed neural representations. Such insight will require *mechanistic* modeling where neurons are explicitly modeled and connected in networks. Starting with the seminal work of Hodgkin and Huxley who developed a mechanistic model for action-potential generation and propagation in squid giant axons (Hodgkin and Huxley, 1952), biophysics-based modeling of neurons is now well established (Koch, 1999; Dayan and Abbott, 2001; Sterratt et al., 2011). Numerous mechanistic neuron models tailored to model specific neuron types have been constructed, for example, for cells in mammalian sensory cortex (Hay et al., 2011; Markram et al., 2015; Pozzorini et al., 2015), hippocampus (Migliore et al., 1995) and thalamus (McCormick and Huguenard, 1992; Halnes et al., 2011).

At the level of networks, most mechanistic studies have focused on generic properties and have considered stylized models with a single or a handful of neuronal populations consisting of identical neurons with statistically identical connection properties. Such studies have given invaluable qualitative insights into the wide range of possible network dynamics (see Brunel (2000) for an excellent example), but real brain networks have heterogeneous neural populations and more structured synaptic connections. For small networks, excellent models aiming to mimic real neural networks have been developed, a prominent example being the circuit in the crustacean stomatogastric nervous system comprising a few tens of neurons (Marder and Goaillard, 2006). However, even though pioneering efforts to construct comprehensive networks with tens of thousands of neurons mimicking cortical columns in mammalian sensory cortices, have been pursued, e.g., Traub et al. (2005); Potjans and Diesmann (2014); Markram et al. (2015); Schmidt et al. (2018a); Arkhipov et al. (2018), mechanistic modelling of biological neural networks mimicking specific brains, or brain areas, is still in its infancy.

A cubic millimetre of cortex contains several tens of thousands of neurons, and until recently, limitations in computer technology have prohibited the mathematical exploration of neural networks mimicking cortical areas even in the smallest mammals. With the advent of modern supercomputers, simulations of networks comprising hundreds of thousands or millions of neurons are becoming feasible. Thus several large-scale brain projects, including the EU Human Brain Project (HBP) and MindScope at the Allen Brain Institute, have endeavoured to create large-scale network models for mathematical exploration of network dynamics (Kandel et al., 2013). In the HBP, where all authors of this paper participate, the goal is not so much to create *models* for specific brain areas, but rather to create general-purpose brain *simulators*. These brain simulators, which also aptly are called brain-simulation engines, will not be tied to specific candidate models but rather be applicable for execution of many candidate models, both current and future candidates. As such their use by the scientific community for mathematical exploration of brain function is expected to go well beyond the planned end of the HBP project in 2023.

In this article, we present the scientific case for brain simulations, in particular the development and use of multi-purpose brain simulators, and argue why such simulators will be indispensable in future neuroscience. Furthermore, the long-term maintenance and continued development of such simulators are not feasible for individual researchers, nor individual research groups. Rather, community efforts as exemplified by the brain-simulator developments in HBP are required.

## 2     Brain simulations

Brain function relies on activity on many spatial scales, from the nanometer scale of atoms and molecules to the meter scale of whole organisms, see, e.g., Devor et al. (2013). And unlike, for example, in a canister of gas, these scales are intimately connected. While the replacement of a single gas molecule with another has no effect on the overall behaviour of the gas, a change in a DNA molecule can change the brain dramatically, like in Huntington's disease (Gusella et al., 1983). Mechanistic models can act as `bridges between different levels of understanding' (Dayan and Abbott, 2001, Preface) as for example in the Hodgkin-Huxley model where axonal action-potential propagation is explained in terms of the properties of ion channels, that is, molecules (proteins) embedded in the cell membrane. Today's most impressive multiscale simulations are arguably the weather simulations that provide, with increased accuracy year by year, our weather forecasts (Bauer et al., 2015). These physics- and chemistry-based simulations bridge scales from tens of meters to tens of thousands of kilometers, the size of our planet, and are in computational complexity comparable to whole-brain simulations (Koch and Buice, 2015).

### 2.1     Brain network simulations

Large-scale brain-simulation projects have until now predominantly focused on linking the neuron level to the network level, that is, simulating synaptically connected networks of hundreds, thousands or more neurons. One obvious reason is that at present such networks, whose properties presumably lie at the heart of our cognitive abilities, are particularly difficult to understand with qualitative reasoning alone, that is, without the aid of mathematics. Another reason is that starting with the seminal works of Hodgkin and Huxley (1952) and Rall (Segev et al., 1994), we now have a biophysically well-founded scheme for simulating how individual neurons process information, that is, how they integrate synaptic inputs from other neurons and generate action potentials. This scheme is covered in all textbooks in computational neuroscience (see, e.g., Koch (1999); Dayan and Abbott (2001); Sterratt et al. (2011)) and typically also in computational neuroscience courses given at universities. Numerous neuron models are now available for reuse and further development and can be downloaded from databases such as ModelDB ([senselab.med.yale.edu/modeldb/](senselab.med.yale.edu/modeldb/)), the Neocortical Microcircuit Collaboration (NMC) Portal ([bbp.epfl.ch/nmc-portal](bbp.epfl.ch/nmc-portal)), the Brain Observatory at the Allen Brain Institute ([observatory.brain-map.org](observatory.brain-map.org)), and Open Source Brain ([opensourcebrain.org](opensourcebrain.org)). Mathematical models for synaptic function, including synaptic plasticity, have also been developed, and all the necessary building blocks for creating models for networks of neurons are thus available.

Some large-scale network models have been based on morphologically detailed neuron models (Reimann et al., 2013; Markram et al., 2015; Arkhipov et al., 2018), some have used stylized spatially-extended neuron models (Traub et al., 2005; Tomsett et al., 2015; Migliore et al., 2015), some have used point neurons of the integrate-and-fire type (Lumer et al., 1997; Izhikevich and Edelman, 2008; Potjans and Diesmann, 2014; Hagen et al., 2016; van Albada et al., 2018; Schmidt et al., 2018a,b), and some have used firing-rate units representing population activity (Schirner et al., 2018). More biological detail does not by itself mean that the model is more realistic. In fact, point neurons, that is, neuron models where the membrane potential is assumed to be the same across dendrites and soma, have been found to be excellently suited to reproduce experimentally recorded action potentials following current stimulation (Jolivet et al., 2008; Pozzorini et al., 2015). The various neuron models have different pros and cons, and the choice of which to use depends on the question asked (Herz et al., 2006). We thus argue that a set of brain simulators for simulation of models at different levels of biological detail should be developed.

For weather simulations the goal is clear, that is, to accurately predict temperature, precipitation and wind at different geographical locations. Likewise, brain simulations should predict what can be experimentally measured, not only action potentials, but also population-level measures such as local field potentials (LFP), electrocorticographic signals (ECoG) and voltage-sensitive dye imaging (VSDI) signals, as well as systems-level measurements such as signals recorded by electroencephalography (EEG) or magnetoencephalography (MEG) (Brette and Destexhe, 2012), cf. Figure 1. For these electrical, magnetic and optical measures the `measurement physics' seems well established, that is, mathematical models for the biophysical link between electrical activity in neurons and what is measured by such recordings have been developed, see references in caption of Figure 1. Simulation tools such as LFPy ([lfpy.github.io](lfpy.github.io)) and BIONET ([alleninstitute.github.io/bmtk/bionet.html](alleninstitute.github.io/bmtk/bionet.html)) for prediction of such electrical and magnetic signals from simulated network activity, both using biophysically-detailed multicompartment models (Lindén et al., 2014; Gratiy et al., 2018; Hagen et al., 2018) and point-neuron models of the integrate-and-fire type (Hagen et al., 2016), are now publically available. For functional magnetic resonance imaging (fMRI) the biophysical link between activity in individual neurons and the recorded BOLD signal is not yet established (but see Uhlirova et al. (2016b,a)), and a mechanistic forward-modeling procedure linking microscopic brain activity to the measurements is not yet available.

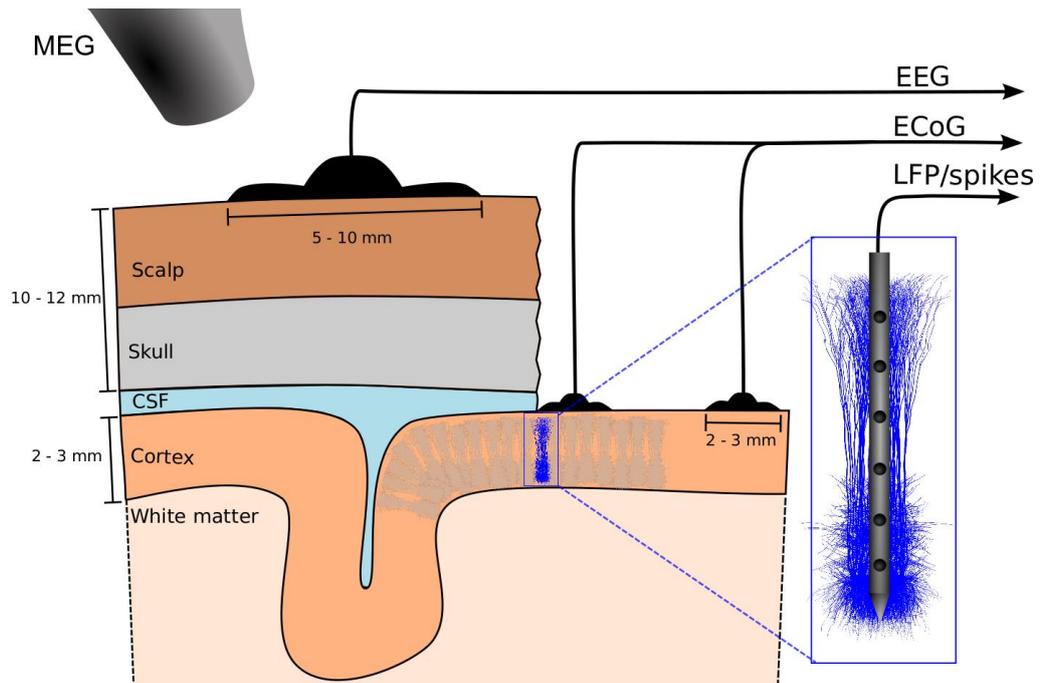

*Figure 1: Electric and magnetic signals to be computed in brain network simulations.*
*Measures of neural activity in cortical populations: (i) spikes (action potentials) and LFP from a linear microelectrode inserted into cortical grey matter, (ii) ECoG from electrodes positioned on the cortical surface, (iii) EEG from electrodes positioned on the scalp, and (iv) MEG measuring magnetic fields stemming from brain activity by means of SQUIDs placed outside the head. For reviews on the biophysical origin and link between neural activity and the signals recorded in the various measurements, see Hämäläinen et al. (1993); Nunez and Srinivasan (2006); Brette and Destexhe (2012); Buzsaki et al. (2012); Einevoll et al. (2013); Pesaran et al. (2018); Hagen et al. (2018).*

Figure 2 illustrates the use of brain network simulators for a so-called barrel column in somatosensory cortex. Each such column primarily processes sensory information from a single whisker on the snouts of rodents, and in rats a barrel column contains some tens of thousands of neurons. A column can be modeled as a network of interconnected neurons based on biophysically-detailed multicompartment models (here referred to as level I), point-neuron models of the integrate-and-fire type (level II), or firing-rate units where each unit represents activity in a neuronal population (level III). Regardless of the underlying neuron type, the simulator should preferably be *multimodal*, that is, simultaneously predict many types of experimental signals stemming from the same underlying network activity. Ideally the neuron models at the different levels of detail should be interconnected in the sense that the simpler neuron models should be possible to reduce from (or at least be compatible with) the more detailed neuron models. The field of statistical physics addresses such scale bridging. A prime example of its application is the development of the thermodynamic ideal-gas law describing the macroscopic properties of gases in terms of variables like temperature or pressure from the microscopic Newtonian dynamics of the individual gas molecules. As a neuroscience example, several projects have aimed to derive firing-rate models (level III) from spiking neuron models (level II), see, e.g., de Kamps et al. (2008); Deco et al. (2008); Ostojic and Brunel (2011); Bos et al. (2016); Schwalger et al. (2017); Heiberg et al. (2018).

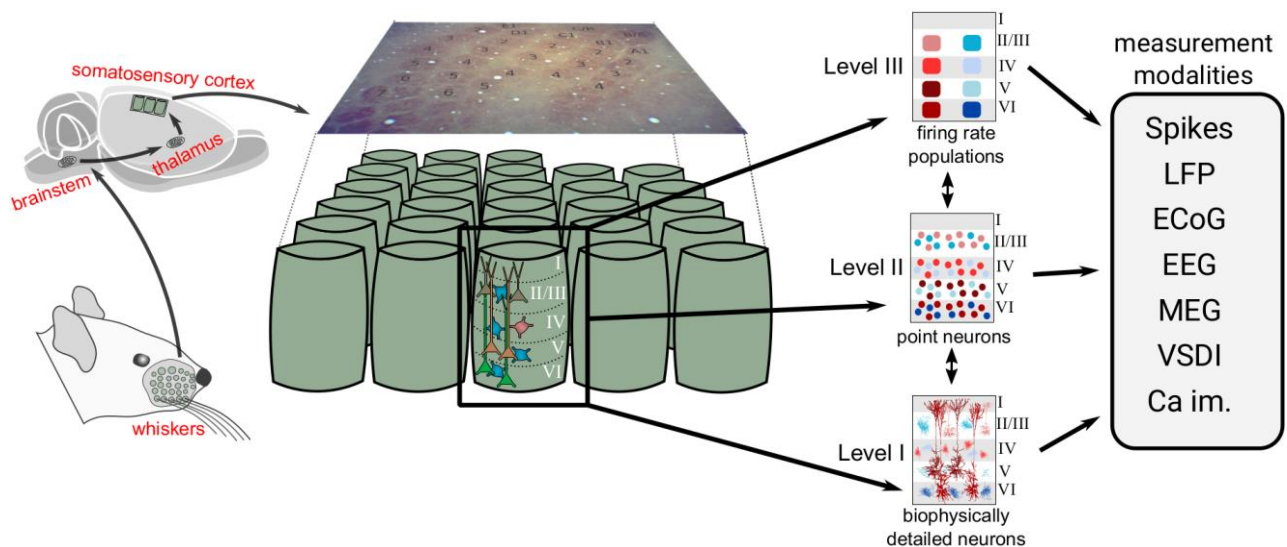

*Figure 2: Illustration of multimodal modeling with brain simulators. Network dynamics in a cortical column (barrel) processing whisker stimulation in rat somatosensory cortex (left) can be modelled with units at different levels of detail. In the present example we have a level organization with biophysically detailed neuron models (level I), simplified point-neuron models (level II), and firing-rate models with neuron populations as fundamental units (level III). Regardless of level, the network simulators should aim to predict the contribution of the network activity to all available measurement modalities. In addition to the electric and magnetic measurement modalities illustrated in Figure 1, the models may also predict optical signals, for example, signals from voltage-sensitive dye imaging (VSDI) signals and two-photon calcium imaging (Ca im.).*

## 2.2 One simulator - many models

When discussing simulations, it is important to distinguish between the model and the simulator. Here

- *model* refers to the equations with all parameters specified,
- *simulator* refers to the software tool that can execute the model (like NEURON, NEST, and The Virtual Brain used in HBP), and
- *simulation* refers to the execution of a model in a simulator.

In some fields of science, simulators are intimately tied to solving a particular model. One example is atomic physics where there is consensus both about what equation to solve (the Schrödinger equation) and the numerical values of the few parameters involved (electron mass, Planck's constant, ...). In contrast, for simulations of brain networks we can and should have a clear separation. The simulators used in HBP are accordingly designed to execute many different models, just like calculus can be used in many different physics calculations. Also, if possible, one should as a control execute the same model on different simulators to check for consistency of the results, see, e.g., van Albada et al. (2018); Shimoura et al. (2018). To facilitate this, software packages for simulator-independent specification of neuronal network models are developed (Davison et al., 2008).

## 3 Network simulators not tailored to specific brain-function hypotheses

In experimental neuroscience the method is often intimately tied to the hypothesis being tested. In electrophysiological experiments, for example, the experimental set-up and execution protocols are tailored to most efficiently answer the biological question asked. In contrast, brain network simulators should not be designed to test a particular hypothesis about brain function, rather they should be designed so that they can test many existing and future hypotheses.

### 3.1 Discovery of Newton's law of gravitation - an analogy

As the tools for testing a hypothesis can easily be confused with the hypothesis itself, we here present an analogy from physics, Isaac Newton's discovery of the law of gravitation. While one could argue that the establishment of a mechanistic understanding of brain function, that is, understanding our cognitive abilities on the basis of neuronal action, would be a breakthrough of similar magnitude as discovering this law, this is not the point here. Here this example is only used to illustrate the role of network simulators in brain research.

Prior to Newton's theory, the planetary orbits had since ancient times been described by the Ptolemaic model. This model assumed the Earth to be at center of the universe and the planets to move in trajectories described by a complicated arrangement of circles within circles (so-called epicycles). The model predicted the planetary orbits accurately and was used for more than 1500 years to make astronomical charts for navigation. It was thus a successful *descriptive* model, but it shed little light on the underlying physical mechanisms governing the planetary movement. As such it had a similarly useful role as the present descriptive, receptive-field-like models accounting for neural representations in the brain.

Newton's theory of gravitation provided a *mechanistic* understanding of planetary movement based on his two hypotheses on (i) how masses attract each other and (ii) how the movement of masses is changed when forces are acting on them. But the theory went beyond planetary orbits in that it, for example, also successfully predicted known kinematic laws of falling apples, trajectories of cannon balls, and high and low tides due to gravitational attraction between the Moon and the water in our oceans.

The first hypothesis of Newton was that two masses $m$ and $M$ with a separation distance $r$ attract each other with a force $F_g$ given by

$$F_g = G \frac{mM}{r^2} , \qquad (1)$$

where $G$ is the gravitational constant. The second hypothesis was that when a force $F$, in this case $F_g$ in Eq. (1), is acting on a mass $m$, the mass will be accelerated with an acceleration $a$ according to

$$a = \frac{F_g}{m}. \qquad (2)$$

To test the validity of these hypotheses, Newton had to compare with experiments, that is, available measurements of planetary orbits. However, the connection between the mathematically formulated hypotheses in Eqs. (1-2) and shapes of predicted planetary orbits is not obvious. In fact, Newton developed a new type of mathematics, calculus, to make testable predictions from his theory to allow for its validation (Leibniz independently developed calculus around the same time). Without the appropriate mathematics it would have been impossible for Newton to test whether nature behave according to his hypotheses.

Comparison with experiments demonstrated that Newtons's hypotheses were correct, and Newton's theory of gravitation is now one of the pillars of physics. However, if it had turned out that the predictions of planetary orbits were not in accordance with the observational data, he could have tried other hypotheses, that is, other mathematically-formulated hypotheses than those in Eqs. (1-2). Then, with the aid of his newly invented calculus, he could have made new planetary orbit predictions and check whether these were in better agreement with measurements.

The point is that calculus was a tool to test Newton's hypotheses about the movement of masses, it was not a part of the hypotheses themselves. Likewise, we argue that brain network simulators should, in analogy to calculus, be designed to be tools for making precise predictions for brain measurements for any candidate hypothesis for how brain networks are designed and operate.

**3.2    Hypothesis underlying brain network simulators**

At present we do not have any well-grounded, and certainly not generally accepted, theory about how networks of millions or billions of neurons work together to provide the salient brain functions in animals or humans. We do not even have a well-established model for how neurons in primary visual cortex of mammals work together to form the intriguing neuronal representations with, for example, orientation selectivity and direction selectivity that were discovered by Hubel and Wiesel sixty years ago (Hubel and Wiesel, 1959). Moreover, we do not have an overview over all neuron types in the brain. However, we do know the biophysical principles for how to model electrical activity in neurons and how neurons integrate synaptic inputs from other neurons and generate action potentials. These principles, which go back to the work of Hodgkin and Huxley (1952) and Rall (Segev et al., 1994) and are described in numerous textbooks (see, e.g., Koch (1999); Dayan and Abbott (2001); Sterratt et al. (2011)), are the only hypotheses underlying the construction of brain network simulators. This is the reason why many models can be represented in the same simulator and why it is possible to develop generally applicable simulators for network neuroscience.

However, while we know the principles for how to model neuronal activity, we do not *a priori* know all the ingredients needed to fully specify network models. In order to construct candidate network models, information on the anatomical structure, electrophysiological properties, and spatial positions of neurons, as well as information on how these neurons are connected, are needed. The MindScope project at the Allen Brain Institute as well as the HBP are gathering such data, and the first large-scale models are constructed on the basis of these and other sources (Arkhipov et al., 2018). Although the primary goal in the HBP is to create general-purpose brain simulators, such initial models are needed to guide the construction of these simulators and to demonstrate their performance and potential usefulness. However, given the present lack of data on, for example, the strength and plasticity of synaptic connections between the neurons, it is clear that these initial models can be nothing more than plausible skeleton models to be used as starting points for further explorations. Experimental data is thus collected to have a starting point for mathematical exploration, not in the belief that brain function will be understood just by collecting these data and 'putting them into a large simulator'.

Each candidate network model with specified neuron models, network structure and synaptic connections precisely defined by a set of model parameters, can be thought of as a candidate hypothesis. Brain network simulators should be designed to allow for the computation of predictions of relevant experimental measures from any such candidate model (see Figure 2) so that the merit of each model can be assessed by comparison with experiments.

In passing, we note that the use of mathematical simulators has a proud history in neuroscience: The accurate model prediction of the speed and shape of propagating action potentials in the squid giant axon by Hodgkin and Huxley in the early 1950s required the numerical solution of the equations on a hand-operated calculation machine, since the newly installed Cambridge computer was down for six months in 1951 (Hodgkin, 1976).

# 4 Use of brain network simulators

## 4.1 Biological imitation game

When the physicist Richard Feynman died in 1988, a statement on his blackboard read: 'What I cannot create, I do not understand.' In the present context an interpretation of this is that unless we can create mechanistic mathematical models mimicking the behaviour in real brains, our understanding will have to remain limited. An obvious use of brain network simulators is to contribute towards building such models. In particular, the simulators should test candidate network models against experiments so that over time network models improve and get closer to the networks that are realized in real biological systems. This would amount to identifying the models that perform best in the 'biological imitation game' (Koch and Buice, 2015), that is, the models whose predictions best mimic experimental recordings of the same system.

In general a unique winner of such an imitation game will not be found, that is, a specific network model with a specific set of model parameters. Rather, classes of candidate models with similar structures and model parameters will likely do equally well, but as more experiments become available the class of models jointly leading this game will expectedly be reduced in size. At all times the leading models can be considered as the currently most promising hypotheses for how the specific biological network is designed and operates, to be challenged by new experiments and new candidate models.

For Newton it was clear what should be compared: the observed planetary orbits and the corresponding orbits predicted by his theory. In brain science this is less clear. Action potentials are clearly the key carrier of information, but what aspects of the trains of action potentials should be mimicked by brain simulations? Detailed temporal sequence of actions potentials from individual neurons, coefficients of variation, or firing rates of individual neurons (Jolivet et al., 2008; Gutzen et al., 2018)? Or maybe the target only should be the average firing rates of populations of neurons? Likewise, it is unclear what aspects of the LFP or VSDI signals should be compared, the full temporal signals or maybe the power spectral densities? The question of what criteria should be used to select the `best' model cannot be fully settled at present. The answer will also depend, for example, on whether one believes information is coded in firing rates or in the detailed temporal structure of action-potential trains. And maybe realistic behaviour of a robot following motor commands produced by a model network could be one success criterium when such models become available? However, such uncertainties regarding modeling targets should not preclude the initiation of a 'biological imitation game', they only mean that different rules of the game may be considered or that the rules might change over time.

Overarching ideas on how computations are performed by the brain can also inspire candidate models. For example, predictive coding (Rao and Ballard, 1999) has emerged as a contender to the more traditional idea that the brain integrates information from the outside world from feature detectors through predominantly feedforward processes. Instead, predictive coding suggests that the brain is constantly updating hypotheses about the world and predicting sensory information by feedback mechanisms. These two competing ideas could, when instantiated as specific network models, make different predictions about neurophysiological experiments.

## 4.2 Validation of data-analysis methods

Another important application of brain simulations is to create benchmarking data for validation of methods used to analyse experimental data. This approach has already been used to generate benchmarking data for testing of automatic spike-sorting algorithms (Hagen et al., 2015), methods for detecting putative synfire chains (Schrader et al., 2008), as well as for testing of methods used to estimate current-source densities (CSDs) from recorded LFPs (Pettersen et al., 2008). Several statistical methods have been developed for estimating, for example, functional connectivities between neural populations and cortical areas based on population-level and systems-level measures such as LFP, EEG, MEG and fMRI signals (Einevoll et al., 2013; Pesaran et al., 2018). These statistical analysis methods should be validated on `virtual' benchmarking data computed by brain network simulators where the ground-truth neuron and network activity are known (Denker et al., 2012). Even if these model-based benchmarking data do not correspond in detail to any specific biological system, data-analysis methods claimed to be generally applicable should also perform well on these simulated data.

**4.3    Use by wider research community**

Anyone who has tried, knows that learning the calculus needed to derive planetary orbits from Newton's hypotheses is demanding, and for most a formal training in mathematics is required. Likewise, the development of brain simulators requires extensive training in mathematics, computer science and physics, as well as a significant coordinated work effort involving many developers. Fortunately, just like a practicing neuroscientist does not need to construct, say, a confocal microscope in order to use it in research, simulators can be used without knowing all the technical inner workings. Some simulators like NEURON (Carnevale and Hines, 2006) even come with a graphical user interface, and plug-and-play programs where neural networks can be created and simulated by pulling elements with the finger onto a canvas, have also been made (Dragly et al., 2017).

While the complexity of many neural network models will make the use of solely graphical user interfaces difficult, it should nevertheless be a goal to design brain simulators so that they also can be used by the general neuroscience community. One aspect of this is that the developers of widely used simulators should regularly offer tutorial and training courses, for example, in connection with major neuroscience conferences. Further, high-quality user-level documentation and support systems for personal inquiries by users must be set up. However, even the best user-level documentation will not enable the general neuroscience community to easily set up large-scale network models. Extracting all necessary neurobiological data from experiments, literature and databases and specifying reliable executable model descriptions, generally require many years of effort as exemplified by recent studies, e.g., Potjans and Diesmann (2014); Markram et al. (2015); Schmidt et al. (2018a); Arkhipov et al. (2018). It is thus essential that executable descriptions of such models are made publicly available as examples and starting points for the community.

Note that the publishing of executable model descriptions may require procedures and tools that go beyond standard scientific publishing practices. For example, the recent publication of a comprehensive multi-area model of macaque visual cortex (Schmidt et al., 2018a,b) was accompanied by detailed model descriptions expressed by technologies like GitHub (https://inm-6.github.io/multi-area-model/) and Snakemake (Köster and Rahmann, 2012) and accompanied by an introductory video (https://youtu.be/YsH3BcyZBcU). Further, the authors also provided the digitized workflow leading from the underlying experimental data to the model specification.

With a plausible candidate network model for, say, a part of V1 in a mouse as a starting point, scientists with modest training in mathematics, physics, and computer science should be able to use brain simulations to ask questions like:

- What is the predicted spiking response of various neuron types to different types of visual stimuli?
- What is the predicted effect on network activity with pharmacological blocking of a particular ion channel in a particular neuron type?
- What does the visually-evoked LFP signal recorded inside the cortex look like, and what neuron populations are predicted to contribute most to this signal?

- Can the EEG signal recorded by electrodes on the scalp positioned outside the visual cortex distinguish between two candidate network models for V1?

The application of brain simulators can be computationally demanding and require use of supercomputers, especially if many model parameter combinations are to be investigated. Not everyone has access to such supercomputers, nor the experience to install or maintain large computer programs. One option would then be to make brain network simulators available through web-based services so that all computations are done remotely on centralized supercomputer centers, as is the plan for HBP.

In the long run, network neuroscience can only approach a mechanistic systems-level understanding if we overcome the complexity barrier by learning how to build on the work of others, that is, by eventually combining models of smaller brain networks components to larger structures of more relevance for cognition. Newton said that he had seen further than others because he was "standing on the shoulders of giants". Likewise, we here argue that we need to find a way to "standing on the shoulders of each other's mathematical models" to have a hope for a detailed understanding of the functioning of brain networks.

# 5      Discussion and outlook

We have here presented arguments for why brain network simulators are not only useful, but likely also critical for advancing systems neuroscience. By drawing the analogy to Newton's discovery of the law of gravity, we have argued that brain simulators should not be made to test specific hypotheses about brain function. Rather, like Newton's development of calculus to allow for testing of the validity of his physical hypotheses regarding planetary movement (Eqs. 1-2), brain simulators should be viewed as `mathematical observatories' to test various candidate hypotheses. A brain simulator is thus a tool, not a hypothesis, and can as such be likened to tools used to image brain structure or brain activity.

In computational neuroscience one has to 'learn to compute without knowing all the numbers' (as quoted from talk by John Hopfield at conference in Sigtuna, Sweden some twenty years ago). What is meant by this is that unlike in, say, quantum-mechanical computations of atomic properties where the handful of model parameters are known to many digits, the model parameters specifying brain networks are numerous, uncertain and may also change over time. The effects of uncertain model parameters and the uncertainty of model predictions can be systematically studied, though such uncertainty quantification requires repeated evaluations of the model of interest and is typically computationally demanding (Tennøe et al., 2018).

We need a set of different brain network simulators describing the neurons at different levels of resolution, that is, different levels of biological detail as exemplified by the three levels depicted in Figure 2. Simulators based on biophysically detailed, multicompartmental neuron models (level I) can explore in detail how the dendritic structures affect integration of synaptic inputs and consequently the network dynamics (Reimann et al., 2013; Markram et al., 2015; Migliore et al., 2015; Arkhipov et al., 2018). Simulators based on point neurons of the integrate-and-fire type (level II) are much less computationally demanding so that larger networks can be studied (Potjans and Diesmann, 2014). Further, the number of model parameters is much smaller and the fitting of single-neuron models to experimental recordings easier (Pozzorini et al., 2015). Population firing-rate models (level III) model the dynamics of entire populations which makes the models computationally, and often also conceptually, much easier (de Kamps et al., 2008; Cain et al., 2016; Schwalger et al., 2017). With population firing-rate models describing a small patch of cortex, so-called neural-mass models, one can derive spatially extended models, so-called neural-field models, covering cortical areas and even complete human cortices (Deco et al., 2008; Ritter et al., 2013; Sanz Leon et al., 2013; Breakspear, 2017; Schirner et al., 2018). While at present most neural-field models are based on largely phenomenological neural-mass models (Jansen and Rit, 1995; Deco et al., 2008), the future goal should be to derive such neural-mass building blocks from population network models based on individual neurons (Zerlaut et al., 2018), or from fitting to experiments (Blomquist et al., 2009).

Brain network simulation is still in its infancy, and the simulators and the associated infrastructure should be developed to allow for the study of larger networks and fully exploit the capabilities of modern computer hardware (see, e.g., Akar et al. (2019) and Kumbhar et al (2019)). They should also allow for the study of longer time-scale processes such as homeostatic and synaptic plasticity (Turrigiano and Nelson, 2004; Keck et al., 2017). With plausible biophysics-based rules for, for example, spike-timing-dependent long-term synaptic plasticity included in the models, studies of learning will also be possible. Brain simulators should eventually also be extended to go beyond the modeling of networks of neurons alone to also incorporate extracellular space and interaction with glia cells (Solbrå et al., 2018). Likewise, they should allow for studies of effects of electrical or magnetic stimulation of the brain, either with intracranial electrodes like in deep-brain stimulation (Perlmutter and Mink, 2006), with surface electrodes (Bosking et al., 2017), or transcranially (Wassermann et al., 2008).

The present paper has focused on brain simulators for studying networks of neurons. While not addressed here, there is clearly also a need for simulators for studying brain activity at the subcellular scale, both for modelling molecular signaling pathways governed by reaction-diffusion dynamics (Bhalla and Wils, 2010) and for modeling molecular dynamics by Newtonian mechanics (Rapaport, 2004). We have also focused on the bottom-up-type network models typically pursued in the computational neuroscience community where model predictions can be compared directly with physiological experiments. However, network models can also be very useful for concisely and precisely representing ideas on how the brain may implement cognitive processes. An early example of such work is the so-called Hopfield model describing how associative memory can be achieved in recurrent networks of binary neurons (Hopfield, 1982). Over the last decades such modeling work has, for example, grown to include visual attention (Reynolds and Desimone, 1999; Deco and Rolls, 2004), language representation (van der Velde and de Kamps, 2006), decision making (Gold and Shadlen, 2007), and learning (Brader et al., 2007). While none of these works immediately predict specific detailed outcomes of neurophysiological experiments, they state ideas about cognitive phenomenon in a concise manner that allows scrutiny and critique.

The development of high-quality brain simulators requires long-term commitment of resources. Both NEURON and NEST, two key brain simulators in the Human Brain Project, have been developed over a time period of more than 25 years. Likewise, the continued development, maintenance and user support of key brain simulators used by the research community will require long-term funding. These simulation tools can be likened to other joint research infrastructures such as astronomical observatories or joint international facilities for studies of subatomic particles. While the expenses for the operation of brain simulators will be much smaller than these experimental facilities, they should nevertheless be considered as necessary research infrastructure and preferably be funded as such.

Until we learn how the wide range of spatial and temporal scales involved in brain function are connected, our understanding of our brains will be limited. Bridging these scales with mathematical modeling will be a daunting challenge, but encouragingly there are examples from other branches of science where many scales have been bridged, the most visible likely being numerical weather prediction (Bauer et al., 2015). Another impressive example of scale bridging is the engineering underlying smart phones. Here tailored materials made of selected semiconductor and metal atoms are assembled into numerous transistors (in some sense analogous to neurons) connected in networks on a chip (`brain'), which together with other components make up the smart phone (`organism'). These examples have been totally dependent on mathematics and simulations to bridge models at different scales. So a natural question is: Do we have a chance of ever understanding brain function without brain simulations?

# Acknowledgments

Funding was received from the European Union Seventh Framework Program (FP7/20072013) under grant agreement No. 604102 (HBP), the European Union's Horizon 2020 Framework Programme for Research and Innovation under Grant Agreements No. 720270 (Human Brain Project SGA1), No. 785907 (Human Brain Project SGA2) and No. 754304 (DEEP-EST), the Research Council of Norway (DigiBrain 248828, CoBra 250128), the Deutsche Forschungsgemeinschaft Grants GR 1753/4-2 and DE 2175/2-1 of the Priority Program (SPP 1665), the Helmholtz Association Initiative and Networking Fund under project numbers ZT-I-0003 (HAF) and SO-902 (ACA), RTG 2416 'Multi-senses Multi-scales', VSR computation time grant Brain-scale simulations JINB33, and the Swiss ETH Domain for the Blue Brain Project.

# References

Akar, A.N., Cumming, B., Karakasis, V., Küsters, A., Klijn, W., Peyser, A., Yates, S. (2019). Arbor – a morphologically-detailed neural network simulation library for contemporary high-performance computing architectures, arXiv e-prints, https://arxiv.org/pdf/1901.07454.pdf

Arkhipov, A., Gouwens, N. W., Billeh, Y. N., Gratiy, S., Iyer, R., Wei, Z., Xu, Z., Abbasi-Asl, R., Berg, J., Buice, M., et al. (2018). Visual physiology of the layer 4 cortical circuit in silico. *PLoS Comput. Biol.* 14, e1006535. doi:10.1371/journal.pcbi.1006535

Bauer, P., Thorpe, A., and Brunet, G. (2015). The quiet revolution of numerical weather prediction. *Nature* 525, 47–55. doi:10.1038/nature14956

Bhalla, U. S. and Wils, S. (2010). *Computational modeling methods for neuroscientists* (MIT Press), chap. Reaction-Diffusion Modeling. 61–92

Blomquist, P., Devor, A., Indahl, U. G., Ulbert, I., Einevoll, G. T., and Dale, A. M. (2009). Estimation of thalamocortical and intracortical network models from joint thalamic single electrode and cortical laminar-electrode recordings in the rat barrel system. *PLoS Comput. Biol.* 5, e1000328. doi:10.1371/journal.pcbi.1000328

Bos, H., Diesmann, M., and Helias, M. (2016). Identifying anatomical origins of coexisting oscillations in the cortical microcircuit. *PLoS Comput. Biol*. 12, e1005132. doi:10.1371/journal.pcbi.1005132

Bosking, W. H., Sun, P., Ozker, M., Pei, X., Foster, B. L., Beauchamp, M. S., and Yoshor, D. (2017). Saturation in phosphene size with increasing current levels delivered to human visual cortex. *J. Neurosci*. 37, 7188–7197. doi:10.1523/JNEUROSCI.2896-16.2017

Brader, J. M., Senn, W., and Fusi, S. (2007). Learning real-world stimuli in a neural network with spike-driven synaptic dynamics. *Neural Comput.* 19, 2881–2912. doi:10.1162/neco.2007.19.11.2881

Breakspear, M. (2017). Dynamic models of large-scale brain activity. *Nat. Neurosci*. 20,

340–352. doi:10.1038/nn.4497

Brette, R. and Destexhe, A. (eds.) (2012). *Handbook of Neural Activity Measurement* (Cambridge University Press)

Brunel, N. (2000). Dynamics of sparsely connected networls of excitatory and inhibitory neurons. *Comput. Neurosci*. 8, 183–208. doi:10.1016/S0928-4257(00)01084-6

Buzsaki, G., Anastassiou, C., and Koch, C. (2012). The origin of extracellular fields and currents–EEG, ECoG, LFP and spikes. *Nat. Rev. Neurosci*. 13, 407–420

Cain, N., Iyer, R., Koch, C., and Mihalas, S. (2016). The computational properties of a simplified cortical column model. PLoS Comput. Biol. 12, e1005045. doi:10.1371/journal.pcbi.1005045

Carnevale, N. T. and Hines, M. L. (2006). *The NEURON Book* (Cambridge University Press)

Davison, A. P., Brderle, D., Eppler, J., Kremkow, J., Muller, E., Pecevski, D., Perrinet, L., and Yger, P. (2008). Pynn: A common interface for neuronal network simulators. *Front. Neuroinf.* 2, 11. doi:10.3389/neuro.11.011.2008

Dayan, P. and Abbott, L. (2001). *Theoretical neuroscience* (MIT Press, Cambridge)

de Kamps, M., Baier, V., Drever, J., Dietz, M., Mösenlechner, L., and van der Velde, F. (2008). The state of miind. *Neural networks* 21, 1164–1181. doi:10.1016/j.neunet.2008.07.006

Deco, G., Jirsa, V. K., Robinson, P. A., Breakspear, M., and Friston, K. (2008). The dynamic brain: from spiking neurons to neural masses and cortical fields. *PLoS Comput. Biol.* 4, e1000092. doi:10.1371/journal.pcbi.1000092

Deco, G. and Rolls, E. T. (2004). A neurodynamical cortical model of visual attention and invariant object recognition. *Vision Res*. 44, 621–642

Denker, M., Einevoll, G., Franke, F., Grün, S., Hagen, E., Kerr, J., Nawrot, M., Ness, T. B., Wachtler, T., and Wojcik, D. (2012). *Report from 1st INCF Workshop on Validation of Analysis Methods*. Tech. rep., International Neuroinformatics Coordinating Facility (INCF)

Devor, A., Bandettini, P. A., Boas, D. A., Bower, J. M., Buxton, R. B., Cohen, L. B., Dale, A. M., Einevoll, G. T., Fox, P. T., Franceschini, M. A., et al. (2013). The challenge of connecting the dots in the B.R.A.I.N. *Neuron* 80, 270–274. doi:10.1016/j.neuron.2013.09.008

Dragly, S.-A., Hobbi Mobarhan, M., Våvang Solbrå, A., Tennøe, S., Hafreager, A., Malthe-Sørenssen, A., Fyhn, M., Hafting, T., and Einevoll, G. T. (2017). Neuronify: An educational simulator for neural circuits. *eNeuro* 4. doi:10.1523/ENEURO.0022-17.2017

Einevoll, G., Kayser, C., Logothetis, N., and Panzeri, S. (2013). Modelling and analysis of local field potentials for studying the function of cortical circuits. *Nat. Rev. Neurosci*. 14, 770–785. doi:10.1038/nrn3599

Gold, J. I. and Shadlen, M. N. (2007). The neural basis of decision making. *Annu. Rev. Neurosci.* 30, 535–574. doi:10.1146/annurev.neuro.29.051605.113038

Gratiy, S. L., Billeh, Y. N., Dai, K., Mitelut, C., Feng, D., Gouwens, N. W., Cain, N., Koch, C., Anastassiou, C. A., and Arkhipov, A. (2018). BioNet: A python interface to neuron for modeling large-scale networks. *PloS ONE* 13, e0201630. doi:10.1371/journal.pone.0201630

Gusella, J. F., Wexler, N. S., Conneally, P. M., Naylor, S. L., Anderson, M. A., Tanzi, R. E., Watkins, P. C., Ottina, K., Wallace, M. R., and Sakaguchi, A. Y. (1983). A polymorphic

dna marker genetically linked to huntington's disease. *Nature* 306, 234–238

Gutzen, R., von Papen, M., Trensch, G., Quaglio, P., Grün, S., and Denker, M. (2018). Reproducible neural network simulations: Statistical methods for model validation on the level of network activity data. *Front. Neuroinf.* 12, 90. doi:10.3389/fninf.2018.00090

Hagen, E., Dahmen, D., Stavrinou, M. L., Linden, H., Tetzlaff, T., van Albada, S. J., Grun, S., Diesmann, M., and Einevoll, G. T. (2016). Hybrid scheme for modeling local field potentials from point-neuron networks. *Cereb. Cortex* 26, 4461–4496. doi:10.1093/cercor/bhw237

Hagen, E., Næss, S., Ness, T. V., and Einevoll, G. T. (2018). Multimodal modeling of neural network activity: computing LFP, ECoG, EEG and MEG signals with LFPy 2.0. *Front. Neuroinf.* 12, 92. doi:10.3389/fninf.2018.00092

Hagen, E., Ness, T. V., Khosrowshahi, A., Sørensen, C., Fyhn, M., Hafting, T., Franke, F., and Einevoll, G. T. (2015). ViSAPy: A python tool for biophysics-based generation of virtual spiking activity for evaluation of spike-sorting algorithms. *J. Neurosci. Methods* 45, 182–204

Halnes, G., Augustinaite, S., Heggelund, P., Einevoll, G. T., and Migliore, M. (2011). A Multi-Compartment Model for Interneurons in the Dorsal Lateral Geniculate Nucleus. *PLoS Comput. Biol.* 7. doi:10.1371/journal.pcbi.1002160

Hämäläinen, M., Hari, R., Ilmoniemi, R., Knuutila, J., and Lounasmaa, O. V. (1993). Magnetoencephalography - theory, instrumentation, and applications to noninvasive studies of the working human brain. *Rev. Mod. Phys*. 65, 413–497

Hay, E., Hill, S., Schürmann, F., Markram, H., and Segev, I. (2011). Models of Neocortical Layer 5b Pyramidal Cells Capturing a Wide Range of Dendritic and Perisomatic Active Properties. *PLoS Comput. Biol*. 7, e1002107. doi:10.1371/journal.pcbi.1002107

Heiberg, T., Kriener, B., Tetzlaff, T., Einevoll, G. T., and Plesser, H. E. (2018). Firing-rate models for neurons with a broad repertoire of spiking behaviors. *J. Comput. Neurosci*. doi:10.1007/s10827-018-0693-9

Herz, A. V. M., Gollisch, T., Machens, C. K., and Jaeger, D. (2006). Modeling single-neuron dynamics and computations: a balance of detail and abstraction. *Science* 314, 80–85. doi:10.1126/science.1127240

Hodgkin, A. L. (1976). Chance and design in electrophysiology: an informal account of certain experiments on nerve carried out between 1934 and 1952. *J. Physiol*. 263, 1–21

Hodgkin, A. L. and Huxley, A. F. (1952). A quantitative description of membrane current and its application to conduction and excitation in nerve. *J. Physiol*. 117, 500–544

Hopfield, J. J. (1982). Neural networks and physical systems with emergent collective computational abilities. *PNAS* 79, 2554–2558

Hubel, D. H. and Wiesel, T. N. (1959). Receptive fields of single neurones in the cat's striate cortex. *J. Physiol*. 148, 574–591

Hyman, S. E. (2012). Revolution stalled. *Sci. Transl. Med*. 4, 155. doi:10.1126/scitranslmed.3003142

Izhikevich, E. M. and Edelman, G. M. (2008). Large-scale model of mammalian thalamocortical systems. *PNAS* 105, 3593–8. doi:10.1073/pnas.0712231105

Jansen, B. H. and Rit, V. G. (1995). Electroencephalogram and visual evoked potential

generation in a mathematical model of coupled cortical columns. *Biol. Cybern*. 73, 357–366

Jolivet, R., Schürmann, F., Berger, T. K., Naud, R., Gerstner, W., and Roth, A. (2008). The quantitative single-neuron modeling competition. *Biol. Cybern.* 99, 417–426. doi: 10.1007/s00422-008-0261-x

Kandel, E. R., Markram, H., Matthews, P. M., Yuste, R., and Koch, C. (2013). Neuroscience thinks big (and collaboratively). *Nat. Rev. Neurosci*. 14, 659–664. doi:10.1038/nrn3578

Keck, T., Toyoizumi, T., Chen, L., Doiron, B., Feldman, D. E., Fox, K., Gerstner, W., Haydon, P. G., Hübener, M., Lee, H.-K., et al. (2017). Integrating hebbian and homeostatic plasticity: the current state of the field and future research directions. *Phil. Trans. R. Soc*. B 372. doi:10.1098/rstb.2016.0158

Koch, C. (1999). *Biophysics of Computation* (Oxford Univ Press, Oxford)

Koch, C. and Buice, M. A. (2015). A biological imitation game. *Cell* 163, 277–280. doi: 10.1016/j.cell.2015.09.045

Köster, J. and Rahmann, S. (2012). Snakemake–a scalable bioinformatics workflow engine. *Bioinformatics* 28, 2520–2522. doi:10.1093/bioinformatics/bts480

Kumbhar, P., Hines M., Fouriaux, J., Ovcharenko, A., King, J., Delalondre, F., Schürmann, F. (2019). CoreNEURON - An Optimized Compute Engine for the NEURON Simulator, arXiv e-prints, https://arxiv.org/pdf/1901.10975.pdf

Lindén, H., Hagen, E., Łęski, S., Norheim, E. S., Pettersen, K. H., and Einevoll, G. T. (2014). LFPy: A tool for biophysical simulation of extracellular potentials generated by detailed model neurons. *Front Neuroinf*. 7. doi:10.3389/fninf.2013.00041

Lumer, E. D., Edelman, G. M., and Tononi, G. (1997). Neural dynamics in a model of the thalamocortical system. I. Layers, loops and the emergence of fast synchronous rhythms. *Cereb. Cortex* 7, 207–227

Marder, E. and Goaillard, J.-M. (2006). Variability, compensation and homeostasis in neuron and network function. *Nat. Rev. Neurosci*. 7, 563–574. doi:10.1038/nrn1949

Markram, H., Muller, E., Ramaswamy, S., Reimann, M. W., Abdellah, M., Sanchez, C. A., Ailamaki, A., Alonso-Nanclares, L., Antille, N., Arsever, S., et al. (2015). Reconstruction and simulation of neocortical microcircuitry. *Cell* 163, 456–492. doi:10.1016/j.cell.2015.09.029

Masters, C. L., Bateman, R., Blennow, K., Rowe, C. C., Sperling, R. A., and Cummings, J. L. (2015). Alzheimer's disease. *Nat Rev. Dis. Primers* 1, 15056. doi:10.1038/nrdp.2015.56

McCormick, D. A. and Huguenard, J. R. (1992). A model of the electrophysiological properties of thalamocortical relay neurons. *J. Neurophysiol*. 68, 1384–1400

Migliore, M., Cavarretta, F., Marasco, A., Tulumello, E., Hines, M. L., and Shepherd, G. M. (2015). Synaptic clusters function as odor operators in the olfactory bulb. *PNAS*, 201502513

Migliore, M., Cook, E. P., Jaffe, D. B., Turner, D. A., and Johnston, D. (1995). Computer simulations of morphologically reconstructed ca3 hippocampal neurons. *J. Neurophysiol.* 73, 1157–1168. doi:10.1152/jn.1995.73.3.1157

Nunez, P. L. and Srinivasan, R. (2006). *Electric fields of the brain: The Neurophysics of EEG* (Oxford University Press, Inc.), 2nd ed.

Ostojic, S. and Brunel, N. (2011). From spiking neuron models to linear-nonlinear models. *PLoS Comput. Biol*. 7, e1001056. doi:10.1371/journal.pcbi.1001056

Perlmutter, J. S. and Mink, J. W. (2006). Deep brain stimulation. *Annu. Rev. Neurosci*. 29, 229–257. doi:10.1146/annurev.neuro.29.051605.112824

Pesaran, B., Vinck, M., Einevoll, G. T., Sirota, A., Fries, P., Siegel, M., Truccolo, W., Schroeder, C. E., and Srinivasan, R. (2018). Investigating large-scale brain dynamics using field potential recordings: analysis and interpretation. *Nat. Neurosci*. 21, 903–919. doi:10.1038/s41593-018-0171-8

Pettersen, K. H., Hagen, E., and Einevoll, G. T. (2008). Estimation of population firing rates and current source densities from laminar electrode recordings. *J. Comput. Neurosci*. 24, 291–313. doi:10.1007/s10827-007-0056-4

Potjans, T. C. and Diesmann, M. (2014). The cell-type specific cortical microcircuit: relating structure and activity in a full-scale spiking network model. *Cereb. Cortex* 24, 785–806. doi:10.1093/cercor/bhs358

Pozzorini, C., Mensi, S., Hagens, O., Naud, R., Koch, C., and Gerstner, W. (2015). Automated high-throughput characterisation of single neurons by means of simplified spiking models. *PLoS Comput. Biol.* 11, e1004275. doi:10.1371/journal.pcbi.1004275

Rao, R. P. and Ballard, D. H. (1999). Predictive coding in the visual cortex: a functional interpretation of some extra-classical receptive-field effects. *Nat. Neurosci*. 2, 79–87. doi:10.1038/4580

Rapaport, D. C. (2004). *The art of molecular dynamics simulation* (Cambridge University Press)

Reimann, M. W., Anastassiou, C. A., Perin, R., Hill, S. L., Markram, H., and Koch, C. (2013). A biophysically detailed model of neocortical local field potentials predicts the critical role of active membrane currents. *Neuron* 79, 375–390. doi:10.1016/j.neuron.2013.05.023

Reynolds, J. H. and Desimone, R. (1999). The role of neural mechanisms of attention in solving the binding problem. *Neuron* 24, 19–29, 111–25

Ritter, P., Schirner, M., McIntosh, A. R., and Jirsa, V. K. (2013). The virtual brain integrates computational modeling and multimodal neuroimaging. *Brain Connect*. 3, 121–145. doi:10.1089/brain.2012.0120

Sanz Leon, P., Knock, S. A., Woodman, M. M., Domide, L., Mersmann, J., McIntosh, A. R., and Jirsa, V. (2013). The virtual brain: a simulator of primate brain network dynamics. *Front. Neuroinf*. 7, 10. doi:10.3389/fninf.2013.00010

Schirner, M., McIntosh, A. R., Jirsa, V., Deco, G., and Ritter, P. (2018). Inferring multi-scale neural mechanisms with brain network modelling. *eLife* 7. doi:10.7554/eLife.28927

Schmidt, M., Bakker, R., Shen, K., Bezgin, G., Diesmann, M., and van Albada, S. J. (2018a). Multi-scale account of the network structure of macaque visual cortex. *Brain Struct. Funct.* 223, 1409–1435. doi:10.1007/s00429-017-1554-4

Schmidt, M., Bakker, R., Shen, K., Bezgin, G., Diesmann, M., and van Albada, S. J. (2018b). A multi-scale layer-resolved spiking network model of resting-state dynamics in macaque visual cortical areas. *PLoS Comput. Biol*. 14, e1006359. doi:10.1371/journal.pcbi.1006359

Schrader, S., Grün, S., Diesmann, M., and Gerstein, G. L. (2008). Detecting synfire chain

activity using massively parallel spike train recording. *J. Neurophysiol*. 100, 2165–2176. doi:10.1152/jn.01245.2007

Schwalger, T., Deger, M., and Gerstner, W. (2017). Towards a theory of cortical columns: From spiking neurons to interacting neural populations of finite size. *PLoS Comput. Biol*. 13, e1005507. doi:10.1371/journal.pcbi.1005507

Segev, I., Rinzel, J., and Shepherd, G. M. (eds.) (1994*). Theoretical Foundations of Dendritic Function: The Collected Papers of Wilfrid Rall with Commentaries* (MIT Press)

Shimoura, R. O., Kamiji, N. L., de Oliveira Pena, R. F., Cordeiro, V. L., Ceballos, C. C., Romaro, C., and Roque, A. C. (2018). Reimplementation of the potjans-diesmann cortical microcircuit model: from nest to brian. *ReScience* 4, 2

Solbrå, A., Bergersen, A., van den Brink, J., Malthe-Sørenssen, A., Einevoll, G., and Halnes, G. (2018). A kirchhoff-nernst-planck framework for modeling large scale extracellular electrodiffusion surrounding morphologically detailed neurons. *PLoS Comput. Biol*. 14, e1006510

Sterratt, D., Graham, B., Gillies, A., and Willshaw, D. (2011). *Principles of computational modelling in neuroscience* (Cambridge University Press)

Tennøe, S., Halnes, G., and Einevoll, G. T. (2018). UncertainPy: A python toolbox for uncertainty quantification and sensitivity analysis in computational neuroscience. *Front. Neuroinf*. 12, 49. doi:10.3389/fninf.2018.00049

Tomsett, R. J., Ainsworth, M., Thiele, A., Sanayei, M., Chen, X., Gieselmann, M. A., Whittington, M. A., Cunningham, M. O., and Kaiser, M. (2015). Virtual electrode recording tool for extracellular potentials (vertex): comparing multi-electrode recordings from simulated and biological mammalian cortical tissue*. Brain Struct. Funct.* 220, 2333–2353. doi:10.1007/s00429-014-0793-x

Traub, R. D., Contreras, D., Cunningham, M. O., Murray, H., LeBeau, F. E. N., Roopun, A., Bibbig, A., Wilent, W. B., Higley, M. J., and Whittington, M. a. (2005). Single-column thalamocortical network model exhibiting gamma oscillations, sleep spindles, and epileptogenic bursts*. J. Neurophysiol*. 93, 2194–232. doi:10.1152/jn.00983.2004

Turrigiano, G. G. and Nelson, S. B. (2004). Homeostatic plasticity in the developing nervous system. *Nat. Rev. Neurosci*. 5, 97–107. doi:10.1038/nrn1327

Uhlirova, H., Kılıç, K., Tian, P., Sakadžic, S., Gagnon, L., Thunemann, M., Desjardins, M., Saisan, P. A., Nizar, K., Yaseen, M. A., et al. (2016a). The roadmap for estimation of cell-type-specific neuronal activity from non-invasive measurements*. Phil. Trans. R. Soc. B* 371, 20150356. doi:10.1098/rstb.2015.0356

Uhlirova, H., Kilic, K., Tian, P., Thunemann, M., Desjardins, M., Saisan, P. A., Sakadzic, S., Ness, T. V., Mateo, C., Cheng, Q., et al. (2016b). Cell type specificity of neurovascular coupling in cerebral cortex. *eLife* 5. doi:10.7554/eLife.14315

van Albada, S. J., Rowley, A. G., Senk, J., Hopkins, M., Schmidt, M., Stokes, A. B., Lester, D. R., Diesmann, M., and Furber, S. B. (2018). Performance comparison of the digital neuromorphic hardware spinnaker and the neural network simulation software nest for a full-scale cortical microcircuit model. *Front. Neurosci.* 12, 291. doi:10.3389/fnins.2018.00291

van der Velde, F. and de Kamps, M. (2006). Neural blackboard architectures of combinatorial structures in cognition. *Behav. Brain Sci*. 29, 37–70; discussion 70–108. doi:10.1017/S0140525X06009022

Wassermann, E., Epstein, C., Ziemann, U., Walsh, V., Paus, T., and Lisanby, S. (2008). *Oxford handbook of transcranial stimulation* (Oxford University Press)

Zerlaut, Y., Chemla, S., Chavane, F., and Destexhe, A. (2018). Modeling mesoscopic cortical dynamics using a mean-field model of conductance-based networks of adaptive exponential integrate-and-fire neurons. *J. Comput. Neurosci*. 44, 45–61. doi:10.1007/s10827-017-0668-2